# Nanoporous $C_3N_4$, $C_3N_5$ and $C_3N_6$ nanosheets; Novel strong semiconductors with low thermal conductivities and appealing optical/electronic properties

Bohayra Mortazavi*[#,a,b], Fazel Shojaei[#,c,d], Masoud Shahrokhi[e], Maryam Azizi[d], Timon Rabczuk[f], Alexander V. Shapeev[g] and Xiaoying Zhuang**[,a,f]

[a]Chair of Computational Science and Simulation Technology, Department of Mathematics and Physics, Leibniz Universität Hannover, Appelstraße 11,30157 Hannover, Germany.
[b]Cluster of Excellence PhoenixD (Photonics, Optics, and Engineering–Innovation Across Disciplines), Gottfried Wilhelm Leibniz Universität Hannover, Hannover, Germany.
[c]Department of Chemistry, College of Sciences, Persian Gulf University, Boushehr 75168, Iran.
[d]School of Nano Science, Institute for Research in Fundamental Sciences (IPM), Tehran, Iran.
[e]Department of Physics, Faculty of Science, University of Kurdistan, Sanandaj, Iran.
[f]College of Civil Engineering, Department of Geotechnical Engineering, Tongji University, Shanghai, China.
[g]Skolkovo Institute of Science and Technology, Skolkovo Innovation Center, Nobel St. 3, Moscow 143026, Russia.

**Abstract**

Carbon nitride two-dimensional (2D) materials are among the most attractive class of nanomaterials, with wide range of application prospects. As a continuous progress, most recently, two novel carbon nitride 2D lattices of $C_3N_5$ and $C_3N_4$ have been successfully experimentally realized. Motivated by these latest accomplishments and also by taking into account the well-known $C_3N_4$ triazine-based graphitic carbon nitride structures, we predicted two novel $C_3N_6$ and $C_3N_4$ counterparts. We then conducted extensive density functional theory simulations to explore the thermal stability, mechanical, electronic and optical properties of these novel nanoporous carbon-nitride nanosheets. According to our results all studied nanosheets were found to exhibit desirable thermal stability and mechanical properties. Non-equilibrium molecular dynamics simulations on the basis of machine learning interatomic potentials predict ultralow thermal conductivities for these novel nanosheets. Electronic structure analyses confirm direct band gap semiconducting electronic character and optical calculations reveal the ability of these novel 2D systems to adsorb visible range of light. Extensive first-principles based results by this study provide a comprehensive vision on the stability, mechanical, electronic and optical responses of $C_3N_4$, $C_3N_5$ and $C_3N_6$ as novel 2D semiconductors and suggest them as promising candidates for the design of advanced nanoelectronics and energy storage/conversion systems.

Corresponding authors: *bohayra.mortazavi@gmail.com; **zhuang@ikm.uni-hannover.de
[#]these authors contributed equally to this work

# 1. Introduction

Two-dimensional (2D) materials are presently among the most attractive and vibrant class of materials, with new members being continuously introduced via theoretical prediction or experimental fabrication. The original interests toward 2D materials was stimulated by the great success of graphene [1–3], which was successfully fabricated in 2004. After the great accomplishments of graphene, carbon-free 2D structures start to gain remarkable attentions. Among various graphene counterparts, transition metal dichalcogenides like $MoS_2$ [4,5], phosphorene [6,7] and indium selenide [8] have been extensively explored during the past decades. One of the main driving forces for the design of aforementioned 2D materials lies in their inherent semiconducting electronic nature, whereas graphene yields zero band-gap electronic character. Presenting a semiconducting electronic character with an appropriate band gap is critical for applications in many advanced technologies, such as: nanoelectronics, nanophotonics, nanosensors, and photocatalysis. While graphene wonderful physics offer numerous possibilities to open a band gap by stretching [9,10], functionalization or doping [11–13] or formation of defects [14–16], but these methods necessitate rather elaborated post fabrication processes, resulting in substantial increase in the production costs. That is why designing inherent 2D semiconductors has been considered as a more viable approach than a band gap opening in graphene. In recent years, a new trend has been evolving and such that tremendous scientific efforts have been devoted to design and fabricate carbon based 2D semiconductors. To achieve this goal, 2D lattices made from carbon and nitrogen covalent networks exhibited remarkable experimental successes. For example, s- and tri-triazine-based graphitic carbon nitride g-$C_3N_4$ [17] nanomembranes with narrow band gaps have been successfully synthesized, and have shown outstanding performances for applications in nanoelectronics, oxygen reduction and photocatalysis [18–22]. Following the remarkable successes of g-$C_3N_4$, in recent years several novel carbon-nitride inherent 2D semiconductors, like $C_2N$ [23], polyaniline $C_3N$ [24], all-triazine $C_3N_3$ [25], triazine $C_3N_5$ [26] and combined triazole/triazine framework of $C_3N_{4.8}$ [27] have been designed and successfully fabricated. These advances highlight the remarkable prospects for the design and fabrication of carbon based 2D semiconductors.

In a recent experimental advance, $C_3N_5$ a novel 2D semiconductor made from two s-heptazine units connected together with azo-linkage was successfully experimentally realized by Kumar *et al.* [28]. According to experimental tests, $C_3N_5$ yields outstanding performances for photocatalytic, photovoltaic and adsorbent applications [28]. In a latest exciting experimental advance, Villalobos *et al.* [29] reported the synthesis of poly(triazine imide), a crystalline graphitic $C_3N_4$, with an electron density gap of 0.34 nm. This novel 2D carbon nitride system was found to exhibit highly desirable performance for high-temperature hydrogen sieving, stemming from its high chemical and thermal robustness [29]. Worthwhile to remind that triazine-based g-$C_3N_4$ show two different atomic lattices, made from s- or tri-triazine groups connected by single N linkages. In this regard, $C_3N_5$ [28] resembles tri-triazine g-$C_3N_4$ in which the connecting single N atom is replaced with N-N azo-linkage. On this basis, we predicted a novel 2D carbon-nitride lattice with a $C_3N_6$ stoichiometry, in which tri-triazine groups in the native $C_3N_5$ are replaced with the s-triazine group, as that of the s-triazine g-$C_3N_4$ [17]. Moreover, $C_3N_4$ lattice fabricated by Villalobos *et al.* [29] is similar to s-triazine g-$C_3N_4$, in which the connecting N atom is terminated by a single H atom. On this basis, we predicted another novel $C_3N_4$ lattice, in which the s-triazine group in the original structure is replaced with the tri-triazine counterpart.

The objective of this study is therefore to explore the thermal stability, mechanical properties, thermal conductivity, electronic and optical characteristics of predicted and most recently experimentally realized 2D carbon nitride systems via first-principles based simulations. To provide a better vision, in our analysis and calculations we also considered the original triazine-based g-$C_3N_4$ nanosheets. In particular, the thermal conductivity was evaluated by employing classical molecular dynamics simulations using machine learning interatomic potentials. The extensive results by this study provide a comprehensive knowledge on the critical properties of four novel carbon nitride 2D systems and will hopefully serve as a valuable guide for the further experimental and theoretical studies.

## 2. Computational methods

Structural optimizations, evaluation of thermal stabilities, electronic structure and optical calculations in this work were performed via density functional theory (DFT) calculations within generalized gradient approximation (GGA) and Perdew–Burke–Ernzerhof (PBE) [30] method. For the DFT calculations we employed *Vienna Ab-initio Simulation Package* (VASP) [31–33]. We used a plane-wave cutoff energy of 500 eV with a convergence criterion of $10^{-4}$ eV for the electronic self-consistent-loop. Periodic boundary conditions were applied along all three Cartesian directions. The simulation box size along the sheets normal direction (z direction) was set to 15 Å, to avoid the interactions with image layers. The geometry optimization for the unit-cells were accomplished using the conjugate gradient method with the a convergence criterion of 0.01 eV/Å for Hellmann-Feynman forces, in which 7×7×1 Monkhorst-Pack [34] k-point mesh size was used. Uniaxial tensile simulations were carried out to evaluate the mechanical properties. Since ordinary DFT within the PBE/GGA underestimate the band gap, we employed hybrid functional of HSE06 [35] to provide more accurate estimations. The thermal stability was examined by conducting ab-initio molecular dynamics (AIMD) simulations at 1000 K and 1500 K with a time step of 1 fs, using a 2×2×1 k-point mesh size.

Absorption coefficient were analyzed on the basis of the random phase approximation (RPA) constructed over the HSE06 results as implemented in the VASP code, with a 500 eV cut-off for the plane wave basis set and a 18×18×1 k-point mesh. The absorption coefficient of the materials is calculated from frequency dependent dielectric constant by using the following relation [36]:

$$\alpha_{\alpha\beta}(\omega) = \frac{2\omega k_{\alpha\beta}(\omega)}{c} = \frac{\omega Im(\varepsilon_{\alpha\beta}(\omega))}{cn_{\alpha\beta}(\omega)} \quad (1)$$

where $c$ is the speed of light in vacuum. $n_{\alpha\beta}$ and $k_{\alpha\beta}$ are real and imaginary parts of the complex refractive index, and are known as the refractive index and the extinction index, respectively. They are given by the following relations:

$$n_{\alpha\beta}(\omega) = \sqrt{\frac{|\varepsilon_{\alpha\beta}(\omega)| + Re(\varepsilon_{\alpha\beta}(\omega))}{2}} \quad (2) \qquad k_{\alpha\beta}(\omega) = \sqrt{\frac{|\varepsilon_{\alpha\beta}(\omega)| - Re(\varepsilon_{\alpha\beta}(\omega))}{2}} \quad (3)$$

Moment tensor potentials (MTPs) [37,38] were employed as an accurate and computationally efficient model of describing interatomic interaction. As discussed in our latest study [39], acquired results for diverse and complex 2D lattices confirm that the trained MTPs can be considered as an alternative to the standard DFT-based method to examine the dynamical stability, phonon dispersion relations, group velocities and other thermal properties. Moreover, MTPs have been recently successfully employed to predict novel materials [38,40], examine lattice dynamics [41,42] and thermal conductivity [43,44]. To create the required training sets, AIMD simulations were carried out at different temperatures of 300, 500, 700 and 900 K, each with 500 time steps. Since the AIMD trajectories are correlated within short intervals, one tenth of subsamples of original trajectories were included in the training sets. Then the MTPs for every particular structure were parameterized by minimizing the difference between the predicted and calculated from the AIMD energies, forces and stresses in the original training sets. After the initial training of MTPs, the accuracy of the trained potentials were evaluated over the full AIMD trajectories and the configurations with high extrapolations grades [45] were selected. The selected configurations were consequently added to the original training sets and the final MTPs were parameterized by retraining of new clean potentials over the updated training sets. The phonon dispersion relations for the predicted t-$HC_3N_4$ and s-$C_3N_6$ monolayers were obtained using the PHONOPY code[46] and the trained MTPs for the force calculations over 6×6×1 supercells.

Non-equilibrium molecular dynamics (NEMD) simulations were carried out to estimate the lattice thermal conductivity of predicted carbon nitride monolayers. To this aim, *Large-scale Atomic/Molecular Massively Parallel Simulator* (LAMMPS) [47] package was used for the NEMD simulations. The trained MTPs were also employed for introducing the interatomic interactions. In our classical molecular dynamics simulations, a time increment of 0.5 fs was used for the full carbon nitride systems, and for those with H atoms a lower time step of 0.25 fs was adjusted to count for the high vibrations of H atoms. To simulate the steady-state heat transfer, we first relax the structures at the room temperature using the Nosé-Hoover thermostat method (NVT). Then few rows of atoms at the two ends were fixed and the rest of simulation box was divided into 22 slabs, as schematically shown in Fig. 1a. Next, 20 K temperature difference was applied between the first (hot) and last (cold) slabs using the NVT method, while the remaining of the system was simulated without applying a thermostat. The applied temperature difference results in the formation of a temperature gradient along the samples (dT/dX) as shown in Fig. 1b. The energy values added (to hot slab) or removed (from cold slab) from the system by the NVT thermostat are close and show linear relations (find Fig. 1c), which can be used to calculate the steady state heat flux, $H_f$. The thermal conductivity was then calculated according to the applied heat flux and established temperature gradient using the one-dimensional form of the Fourier law, assuming a thickness of 3.35 Å as that of the graphene monolayer.

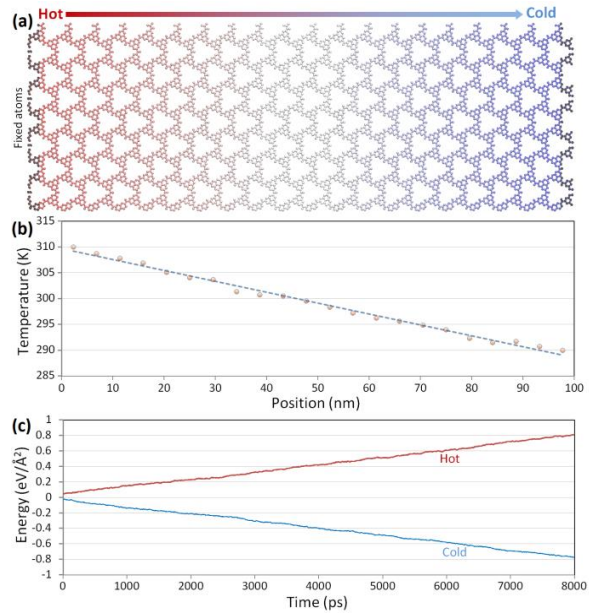

**Fig. 1**, (a) schematic illustration of NEMD simulations to evaluate the lattice thermal conductivity. The atoms at the two ends are fixed and a constant temperature difference is applied between the two ends by the NVT method throughout the simulations. (b) Averaged temperatures in every slab show a linear relation as shown by the dashed line. (c) Applied energies by the NVT method to the hot and cold reservoirs to keep the applied temperature difference constant. The slope of these energy curves can be used to calculate the applied steady state heat flux in the system.

## 3. Results and discussions

We first explore the atomic structures of the considered carbon nitride systems in this work. In Fig. 2, geometry optimized carbon nitride lattices investigated in this study are illustrated. Famous triazine-based g-$C_3N_4$ show two atomic lattices, consisting of single- ($C_3N_3$) or tri- ($C_6N_7$) triazine groups connected by single N atoms, which are illustrated in Fig. 2a and b, respectively. These two structures in their unit cells include $C_3N_4$ and $C_6N_8$, with the hexagonal lattice constants of 4.784 and 7.134 Å, respectively, for their fully flat configurations. The lattice constant of t-$C_3N_4$ has been reported to be 7.13 Å [48], 7.1 Å [49] and 7.15 Å [50] and for s-$C_3N_4$ it was measured to be 4.774 Å [51], 4.74 Å [52], which are close to our results. When comparing different structures in this work, core groups and linkages are the two main features in defining every structure. This way the chemical cores include $C_3N_3$ or $C_6N_7$ groups, illustrated in Fig. 2 by the green and red dotted lines, respectively. The connecting groups include: N, NH and N-N linkages. Similarly when the connecting group is a single N atom, the resulting structures are prominent $C_3N_4$ nanosheets. In Fig. 2c the most recently experimentally realized $C_3N_4$ lattice by Villalobos *et al.* [29] is presented. In accordance with the experimental work, we call this structure as s-H$C_3N_4$, "s" referring to $C_3N_3$ core groups and "H" due to presence of H atoms. Nevertheless, the unit cell of aforementioned structure includes 6, 9 and 3; C, N and H atoms, respectively. In Fig. 2d, "t-$HC_3N_4$" the novel 2D lattice predicted by study is illustrated. This novel nanosheets is basically the same as s-$HC_3N_4$ but made of larger $C_6N_7$ cores, which in the unit cell includes 12, 17 and 3; C, N and H atoms, respectively. The hexagonal lattice constants of s-$HC_3N_4$ and t-$HC_3N_4$ were measured to be 8.647 and 12.972 Å, respectively. Our estimated lattice constant for s-$HC_3N_4$ is remarkably close to the value of 8.645 Å [29] reported in the original experimental work. The $C_3N_5$ lattice realized by Kumar *et al.* [28] is shown in Fig. 2f, which is made of $C_6N_7$ cores connected by N-N linkages. By replacing the $C_6N_7$ cores with $C_3N_3$ counterparts and keeping the same connecting group, another novel carbon nitride 2D system with a $C_3N_6$ stoichiometry form, which is shown in Fig. 2e. The hexagonal lattice constants of $C_3N_6$ and $C_3N_5$ were found to be 10.575 and 15.128 Å, respectively. For the convenience of future studies, geometry optimized lattices in the VASP POSCAR format are included in the supporting information. Although carbon nitride nanosheets are well known for their covalent bonding nature, to further demonstrate this concept the electron localization function (ELF) [53] is also plotted in Fig. 2. ELF is a spatial function taking a value between 0 and 1 for each point of the real space, in which values close to 1 indicate strong covalent bonding or lone pair electrons, and values lower than 0.5 correspond to the electron gas and low electron density localization. As expected, the electron localization occurs around the center of C-N and N-H bonds confirming the dominance of covalent bonding in these systems. Noticeable electron localization also happens around core edge N atoms stemmed from their lone pair electrons.

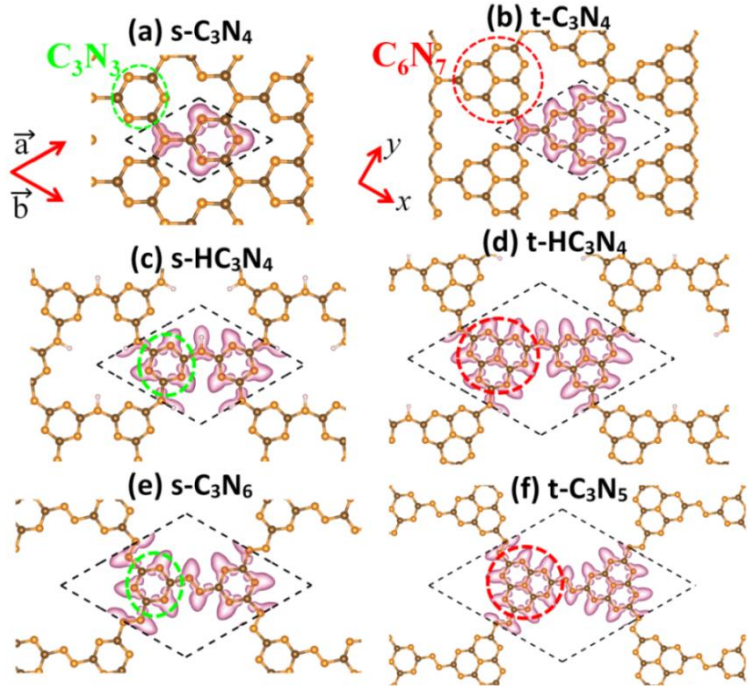

**Fig. 2**, Atomic structure of carbon nitride monolayers studied in this work. Iso-surface illustrates the electron localization function [53] within the unit cell, which is set at 0.7.

First we calculate the elastic constants of s-HC$_3$N$_4$, t-HC$_3$N$_4$, s-C$_3$N$_6$ and t-C$_3$N$_5$ monolayers to examine their elastic stability. According to our results, these considered lattices were found to be isotropic, such that C$_{11}$= C$_{22}$. The C$_{11}$ (C$_{12}$) elastic constants of s-HC$_3$N$_4$, t-HC$_3$N$_4$, s-C$_3$N$_6$ and t-C$_3$N$_5$ were measured to be 59.6 (5.4), 39.3(0.2), 56.0(22.3) and 45.4(18.23) N/m, respectively. The corresponding C$_{66}$ values were also calculated to be 31, 23, 18.5 and 17.2 N/m. For these monolayers, C$_{11}^2$-C$_{12}^2$>0 and C$_{66}$>0 conditions [54] are observed and such that they show elastic stability. In Fig. S1, the acquired phonon dispersion relations for the predicted s-C$_3$N$_6$ and t-HC$_3$N$_4$ monolayers along the high symmetry directions of the first Brillouin zone are illustrated. As the characteristic of 2D materials, these novel nanosheets also show three accoustoic modes starting from the Γ point and none of them exhibit imaginary frequencies, which confirm the dynamical stability. For the case of t-HC$_3$N$_4$ due to the presence of H atoms, flat bands appear in the phonon dispersion relations at frequencies around 104.3 THz, which are not shown in Fig. S1.

Next we explore the mechanical properties of the considered carbon nitride nanomembranes by performing uniaxial tensile results. Since we deal with monolayers that are in contact with vacuum along their normal direction, the stress along the sheet's normal direction remains naturally negligible after the geometry optimization. Nonetheless, for the strained samples along the loading direction, without appropriately adjusting the cell size along the perpendicular direction of loading, most probably the stress component along this direction evolves due to the Poisson's ratio effect. To satisfy the uniaxial stress condition, the box size along the perpendicular direction of loading was accordingly adjusted to reach a negligible stress after the geometry optimization process. In order to evaluate anisotropicity in the mechanical properties, uniaxial tensile simulations were conducted along the two perpendicular planar directions. The predicted uniaxial stress-strain response of studied carbon nitride nanosheets along different loading directions are compared in Fig. 3. Similar to conventional materials, the stress-strain curves illustrate initial linear relations, corresponding to the linear elasticity. The slope of these initial linear relations can be used to extract the elastic modulus. As the first finding it is conspicuous that in all studied porous carbon nitride monolayers, the initial linear relations coincide closely for the loading along $x$ and $y$ directions, revealing convincingly isotropic elasticity. The elastic modulus of s-$C_3N_4$, t-$C_3N_4$, s-$HC_3N_4$, t-$HC_3N_4$, s-$C_3N_6$ and t-$C_3N_5$ were estimated to be 225, 172, 59, 39, 47 and 38 N/m, respectively. The corresponding Poisson's ratios were also measured to be 0.17, 0.24, 0.09, 0.0, 0.40 and 0.41, respectively. As expected, the structures with larger porosities exhibit declined elastic modulus.

Apart from the elastic modulus, another key mechanical property for the engineering design is the maximum tensile strength. Moreover, the corresponding strain at maximum tensile strength is another important mechanical feature, which defines the stretchability of the structure and correlates to its ability to absorb energy before rupture. It is noticeable that apart from the original $C_3N_4$ lattices, for the rest of considered samples the maximum tensile strengths and the corresponding strains are very close for different loading directions. To better understand the origin of anisotropic mechanical responses in the original $C_3N_4$ lattices, one should examine the orientation of bonds during the deformation process. During the deformation process, the bonds oriented along the loading direction elongate and contribute directly to the lead bearing, whereas those oriented along the perpendicular direction of loading contribute marginally in the load transfer. Our analysis of ruptured samples along the both loading directions confirm that in all cases the first debonding in the studied nanosheets occur in the C-N bonds connecting the cores and linkages (find Fig. 3 insets). In other words, these C-N bonds play critical roles on the maximum tensile strength and the corresponding strain. In the original $C_3N_4$ lattices, for the loading along the $y$ direction, these bonds are exactly in-line with the loading, and such that the deformation at every step directly results in the elongation of these bonds. On the other side for the loading along the *x* direction, these CN bonds are initially oriented along the loading with an angle of 30 degree. During the loading these bonds not only elongate but also rotate to orient along the loading direction. Since the elongation of these bonds is limited, the structures fail earlier for the loading along the *y* direction in comparison with those along the *x* direction, in which these critical C-N bonds not only stretch but also rotate during the deformation. For s-$HC_3N_4$, t-$HC_3N_4$, s-$C_3N_6$ and t-$C_3N_5$ the tensile strength values and corresponding strains are very close for the two different loading directions. According to acquired uniaxial stress-strain relations, the maximum tensile strength of s-$C_3N_4$, t-$C_3N_4$, s-$HC_3N_4$, t-$HC_3N_4$, s-$C_3N_6$ and t-$C_3N_5$ were estimated to be 14.4, 9.5, 10.5, 6.7, 4.6 and 3.2 N/m, occurring at corresponding strain levels of 0.13, 0.11, 0.22, 0.2, 0.15 and 0.12, respectively. It is clear that by increasing the porosity the tensile strength and corresponding strain decrease. These results also highlight the substantial role that the type of linkage plays on the maximum tensile strength, in which the structures with NH linkages show more than twice strengths in comparison with corresponding lattices made of N-N counterparts. The results for mechanical properties confirm elastic stability and also reveal remarkable strength and acceptable stretchability of studied nanosheets, stemming from their strong covalent bonding networks. Another critical issue with respect to the stability of a novel material lies in the thermal stability and the ability of lattice to stay intact at high temperatures. We therefore examined the thermal stability by conducting the AIMD simulations at high temperatures of 1000 and 1500 K for 10 ps long simulations. Our analysis of AIMD trajectories confirms that all the considered carbon nitride nanosheets could stay completely intact at the very high temperature of 1500 K (find Fig. S2). Therefore, our results confirm the desirable thermal and mechanical stability of studied nanomembranes.

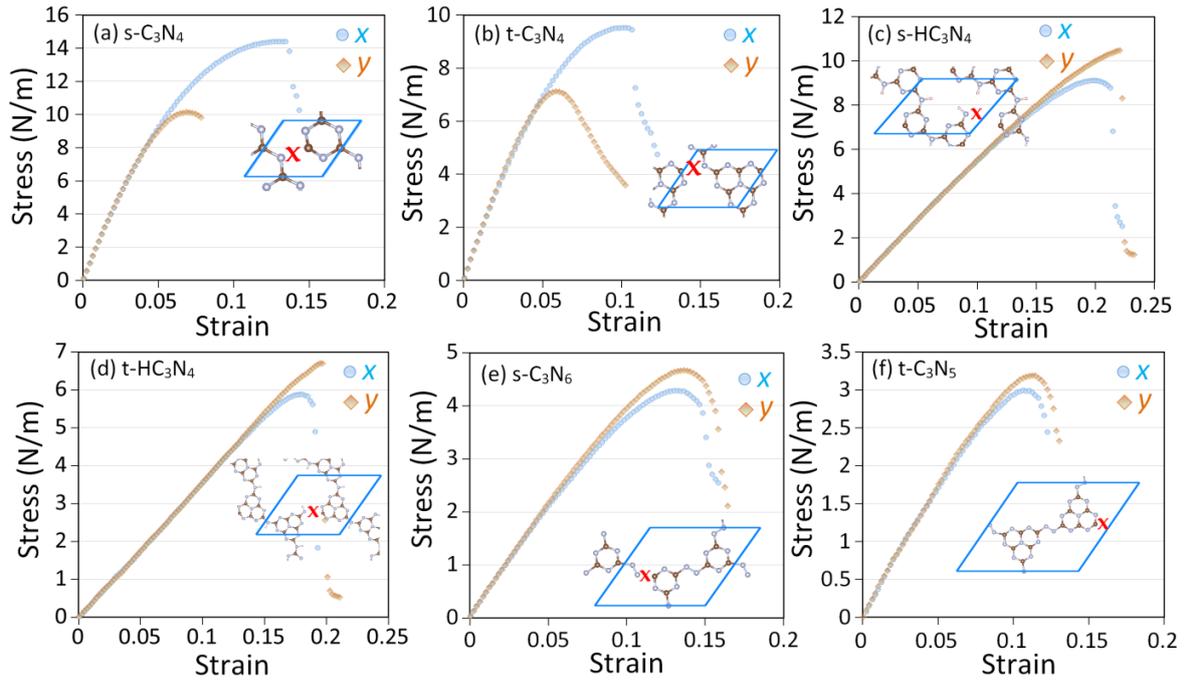

Fig. 3, Uniaxial stress-strain relations of different carbon nitride nanosheets elongated along the *x* and *y* directions. Insets show the structures after the tensile strength point under uniaxial loading along the *x* direction, the corresponding strain levels for every panel are respectively as follows; 0.15, 0.13, 0.24, 0.2, 0.15 and 0.13. The cross sign illustrates the locations that the first debonding occurs.

We next examine the lattice thermal conductivity of t-HC$_3$N$_4$, s-C$_3$N$_6$ and t-C$_3$N$_5$ monolayers on the basis of first-principles based NEMD simulations. Because of the fact that in the NEMD simulations the atoms at the two ends are fixed in order to apply the loading conditions, this may imposes limitation on the contribution of long wavelength phonons. Therefore one has to investigate the length effect on the predicted thermal conductivity to ensure the length indecency of the final estimations. Worthy to note that using the optimized Tersoff potential [55], the converged thermal conductivity of s-C$_3$N$_4$ and t-C$_3$N$_4$ monolayers were recently predicted [56] to be 44 and 17 W/mK, respectively. These are consistent with the corresponding values of 49.8 and 14.1 W/mK, respectively, reported by Dong *et al*. [57]. Nonetheless, using the ReaxFF interatomic potential Dong *et al*. [57] reported distinctly higher thermal conductivities of 111.9 and 50.9 W/mK, respectively, for the single-layer s-C$_3$N$_4$ and t-C$_3$N$_4$. Discrepancy between estimations by different classical interatomic potentials is a common issue/challenge when employing the classical MD simulations [44,58–61]. However, despite quantitative differences these results comparatively show that when the linkage is made of a single N atom, the type of cores strongly affect the thermal conductivity. Now, we consider the thermal transport along the highly porous carbon nitride counterparts. We remind that these NEMD simulations on the basis of MTPs are within, DFT level of accuracy for the computed energies and forces, and such that these results are expected to be more accurate and trustable in comparison with simulations on the basis of classical force fields. The NEMD results for the length effect on the lattice thermal conductivity of t-HC$_3$N$_4$, s-C$_3$N$_6$ and t-C$_3$N$_5$ monolayers at room temperature on the basis of passively trained MTPs are plotted in Fig. 4. As observable for the three studied samples, for small lengths the thermal conductivity shows initial increases by increasing the length. Nonetheless, for the samples with lengths over 50 nm, it is clear that the thermal conductivity converges and reaches the diffusive heat transfer regime. According to our results, the thermal conductivity of t-HC$_3$N$_4$, s-C$_3$N$_6$ and t-C$_3$N$_5$ monolayers at room temperature were estimated to be 5.2± 0.2, 2.5± 0.1 and 2.0± 0.1 W/mK, respectively. Considerably close thermal conductivities of s-C$_3$N$_6$ and t-C$_3$N$_5$ nanosheets, which are almost half of that of t-HC$_3$N$_4$ counterpart, clearly reveals the substantial role of linkages on the lattice thermal conductivity of these highly porous nanostructures. On this basis, the lattice thermal conductivity of s-HC$_3$N$_4$ is expected to be slightly higher than the t-HC$_3$N$_4$ nanosheets. By taking into account the classical and first-principles based results, it can be concluded that for the low porous structures as those in s-C$_3$N$_4$ and t-C$_3$N$_4$, the type of core structures dominate the thermal transport, in contrast with the highly porous counterparts in which the type of linkages dictate the thermal transport. The ultralow thermal conductivity of s-C$_3$N$_6$ and t-C$_3$N$_5$ nanosheets along with their semiconducting electronic nature might be particularly promising to design advanced thermoelectric nanodevices.

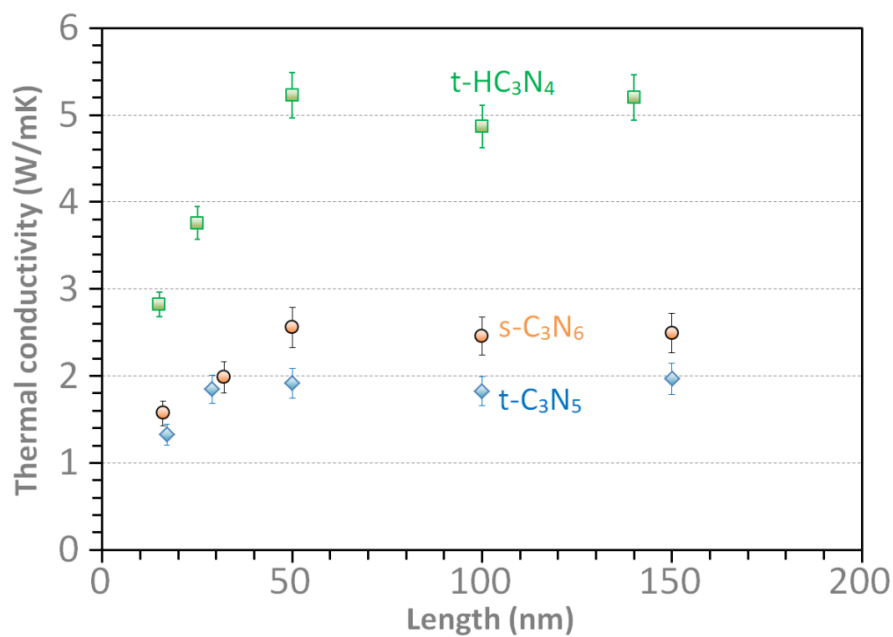

**Fig. 4**, NEMD results for the length effect on the lattice thermal conductivity at room temperature.

Next, we investigate the electronic properties of all these six considered carbon nitride monolayers. The electronic band structures of considered monolayers computed with PBE and hybrid HSE06 functionals, as well as their partial density of states (PDOS) are shown in Fig. 5. We also carefully checked the magnetic properties of these systems by performing spin-polarized calculations. No magnetic moment was observed for any of them, indicating that they all are diamagnetic. According to Fig. 5, the studied monolayers are all semiconductors with HSE06 (PBE/GGA) estimated band gaps of 3.17(1.55) eV for s-$C_3N_4$, 2.80(1.21) eV for t-$C_3N_4$, 4.74(3.25) eV for s-$HC_3N_4$, 3.30(2.33) eV for s-$HC_3N_4$, 2.59(0.89) eV for s-$C_3N_6$, and 2.12(0.63) eV for t-$C_3N_5$. As expected, PBE/GGA predictions are appreciably smaller than their HSE06 counterparts. However, both functionals give almost similar structures and band dispersion patterns. The band gap of s-$C_3N_4$ by the HSE06 (PBE/GGA) were reported to be 3.22 eV [62], 2.97(1.59 eV) [63], 3.2 eV [64] and (1.62 eV [65]). Moreover, the band gap of t-$C_3N_4$ by the HSE06 (PBE/GGA) were reported to be 2.71 eV [66], 2.7 eV [52], 2.7(1.9) eV [67], (1.24 eV [48]) and (1.18 eV [50]), which show close agreements with our results for these two structures. As shown in Fig. 5, these monolayers exhibit direct band gaps at $\Gamma$-point, except for t-$C_3N_4$, in which the valance band maximum (VBM) is located at $\Gamma$-point, while the conduction band minimum (CBM) occurs at K-point. t-$C_3N_4$ has a direct gap of 2.86 eV at M-point, slightly greater (0.06 eV) than its indirect gap (2.80 ($\Gamma\rightarrow K$)). The energy difference between direct and indirect gaps of t-$C_3N_4$ monolayer is comparable to the thermal energy at room temperature (0.026 eV), therefore it can be considered as quasi-direct gap semiconductor. The experimentally measured band gap of multilayered t-$C_3N_5$ system (1.76 eV) [28] is about 0.36 eV smaller than our HSE06 value (2.12 eV) for t-$C_3N_5$ monolayer. This observation can be simply explained by considering the fact that the quantum confinement of charge carriers increases with decreasing the number of layers, leading to the increase of band gap.

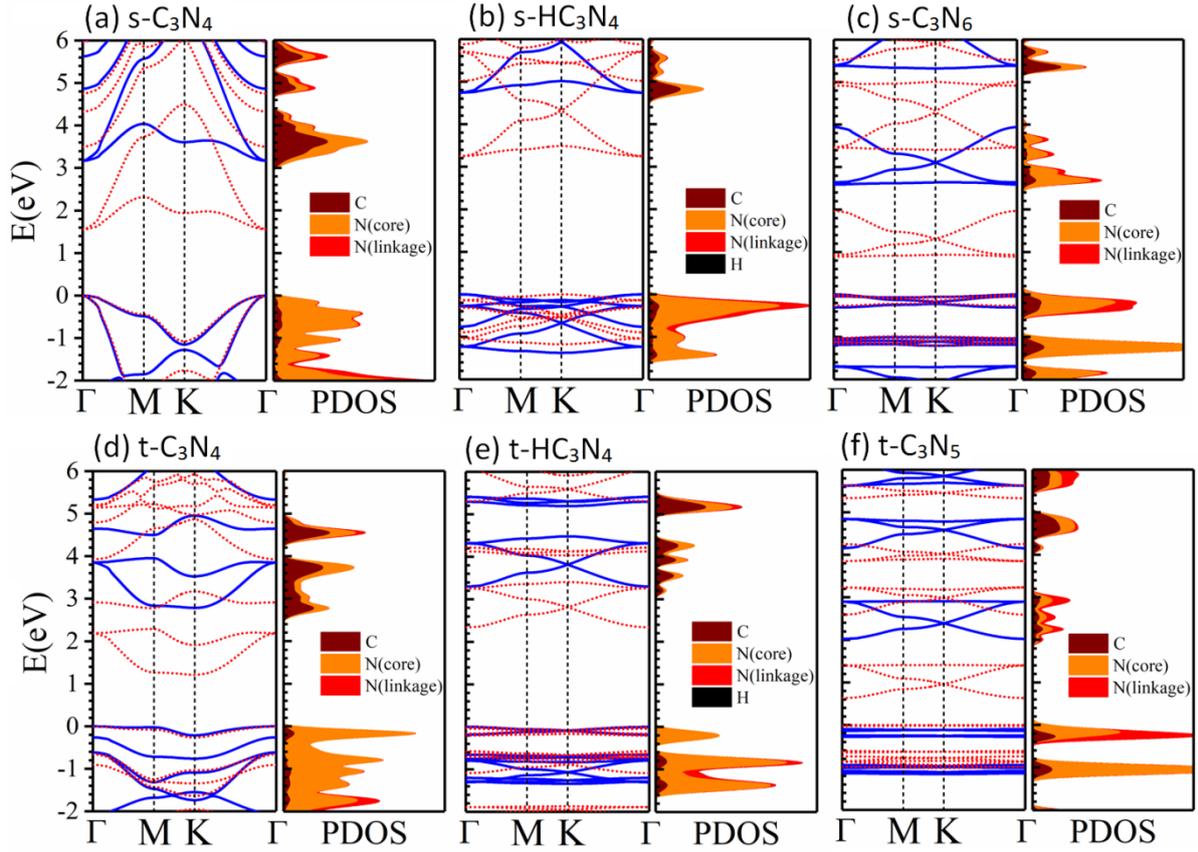

Fig. 5, Electronic band structures and partial density of states (PDOS) of the six carbon nitride monolayers considered in this work, computed using PBE (dotted lines) and hybrid HSE06 (continuous lines) functionals. For both the PBE and HSE06 band structures we set the valance band maximum to zero energy.

Now we take a closer look at the evolution of electronic band structures of these six carbon nitride monolayers as the linkage group changes from tertiary amine (N) to imide (NH) and azo (N=N) groups. To better understand the nature of frontier states, we also calculated the charge density distribution at VBM and CBM of each monolayer (shown in Fig. S3). From Fig. 5, it can be seen that $t-C_3N_4$ monolayer exhibit an almost flat valance band, different with $s-C_3N_4$ counterpart. Such flat bands are associated with very large charge carrier effective masses, low carrier mobilities, and large DOSs in the corresponding energy region. Careful analysis of PDOS and charge density distributions of $s-C_3N_4$ and $t-C_3N_4$ monolayers reveals that for both monolayers, VBM is mainly derived of in-plane lone pair electrons of core nitrogen atoms and also in-plane $p$ orbital carbon atoms, while the CBM seems to be quite different in character and it is almost equally contributed by normal to the plane $p_z$ orbitals of core carbon and nitrogen atoms, representing a $\pi^*$ state (see Fig.5 and Fig. S3). Besides having different band gaps, the band structures of $s-HC_3N_4$ and $t-HC_3N_4$ are found to share many similarities. Both monolayers exhibit almost flat valence bands and dispersive conduction bands, indicating large hole effective masses while small electron effective masses. The large difference between the electron and hole effective masses may help on efficient separation of photo-generated electron-hole pairs. The nature of VBM and CBM of s-HC$_3$N$_4$ and s-HC$_3$N$_4$ are quite similar to those of C$_3$N$_4$ monolayers except for the VBM of t-HC$_3$N$_4$ which is solely made of $p_z$ orbitals of core N atoms (see Fig.5 and Fig. S3). The electronic structures of s-C$_3$N$_6$ and t-C$_3$N$_5$ are different than the other four discussed monolayers. For s-C$_3$N$_6$ both valence band and conduction bands are rather flat. The two states are degenerates at $\Gamma$-point. The conduction sub-band at this point is much more dispersive than valance sub-band. For t-C$_3$N$_5$ the conduction band is found to be dispersive and only valance band is completely flat. This non-trivial flatness has been also observed in some other two dimensional π conjugated COF or metal organic frameworks with a special lattice, called kagome lattice. These kinds of systems can be appealing for strongly correlated states and related magnetic or superconducting properties. Our analysis indicates that for both s-C$_3$N$_6$ and t-C$_3$N$_5$ the VBM is mainly contributed by in-plane p orbitals of N atoms belonging to the azo group and those carbon atoms directly bonded to the linkage atoms, representing lone pair states hybridized with bonding σ(C-N) states. However, the conduction band is made of normal to the plane p$_z$ orbitals of all atoms belong to the core and linkage group, representing a π state (see Fig.5 and Fig. S3).

We found that type of linkage group (N, N=N, NH) yields a strong influence on the electronic properties of the triazine- (s-) and heptazine-based (t-) monolayers. In comparison with the well-known s-C$_3$N$_4$ and t-C$_3$N$_4$ monolayers with tertiary amine linkage groups, s-C$_3$N$_6$ and t-C$_3$N$_5$ monolayers with azo linkage groups have smaller band gaps, while s-HC$_3$N$_4$ and s-HC$_3$N$_4$ monolayers with imide group linkage groups have substantially larger band gaps. In addition, it is also found that for each functional group, triazine-based monolayers have larger band gaps than heptazine ones. These observations can be rationalized by noting that in a 2D material, extending the conjugated π network results in decreasing the band gap, while the gap increases when the conjugated π network is shortened. According to the charge density distribution of VBMs and CBMs of these monolayers, it can be clearly seen that the conjugated π system of a triazine motif (C$_3$N$_3$) is smaller than that of a heptazine (C$_6$N$_7$), serving larger band gaps for triazine-based monolayers. On one hand, as pointed out by Kumar *et al.* [28], replacing the tertiary amine linkage groups with the azo groups extends the conjugated π network due to the overlap between *p* orbitals of N atoms of azo group and π system of the core motif (triazine or heptazine), resulting in lower band gaps for azo-based monolayers. This is evidenced by the significant contribution of p$_z$ orbitals azo N atoms to the conduction states of s-C$_3$N$_6$ and t-C$_3$N$_5$ monolayers, according to their PDOSs and charge density distributions of CBMs. linkage nitrogen atoms to the conduction states of On the other hand, as it can be clearly seen in Fig. 1, s-HC$_3$N$_4$ and s-HC$_3$N$_4$ can be constructed from their C$_3$N$_4$ counterparts by removing one of every seven C$_3$N$_3$ or C$_6$N$_7$ motifs and saturating the N linkers with H atoms. Therefore, s-HC$_3$N$_4$ and s-HC$_3$N$_4$ have much smaller conjugated π network, resulting in substantially larger band gaps.

We next analyze the optical response of these novel 2D nanosheets. Because of the huge depolarization effect in the 2D planar geometry for the out-of-plane light polarization, the optical absorption spectrum are reported only for the in-plane light polarizations [68]. The calculated absorption coefficient $α_{ij}(ω)$ within RPA+HSE06 for in-plane light polarization are plotted in Fig. 6. In this case we also compared the acquired results in the UV-vis window (from 300 to 600 nm) as a function of wavelength. The first absorption peak for s-C$_3$N$_4$ nanosheet occurs at 5.07 eV which is in middle ultraviolet range (MUV, 4.13-6.20 eV) and related to π→π* transitions. It has been experimentally observed and theoretically explained that n→π* electronic transitions are inactive for s-C$_3$N$_4$ monolayer [69–72]. Therefore, the first absorption peak of s-C$_3$N$_4$ monolayer (π→π* electronic transition) occurs at much higher energies with respect to the energy of its band gap transition. As a result, the first absorption peak of all other novel 2D monolayers has a red shift when compared with s-C$_3$N$_4$ nanosheet. Our results show that the first absorption peaks of t-C$_3$N$_4$, t-C$_3$N$_5$ and t-HC$_3$N$_4$ monolayers occur at energy of 3.70, 2.85 and 3.75 eV, respectively, which are desirable for the practical applications in optoelectronic devices in the visible spectral range. The first absorption peaks of s-C$_3$N$_6$ and s-HC$_3$N$_4$were found to locate at the energy of 4.37 and 4.88 eV which are in near ultraviolet (NUV) and MUV range of light. These first absorption peaks are due to electron transition from *p* orbitals of core N atoms (valance band) to *p* orbitals of core carbon and nitrogen atoms (conduction band) (see Fig. 5). In general, the high absorption coefficients were attained (~10$^5$ cm$^{-1}$) for t-C$_3$N$_4$, t-C$_3$N$_5$ and t-HC$_3$N$_4$ monolayers in visible range of light which may be interesting for visible-light optoelectronic applications.

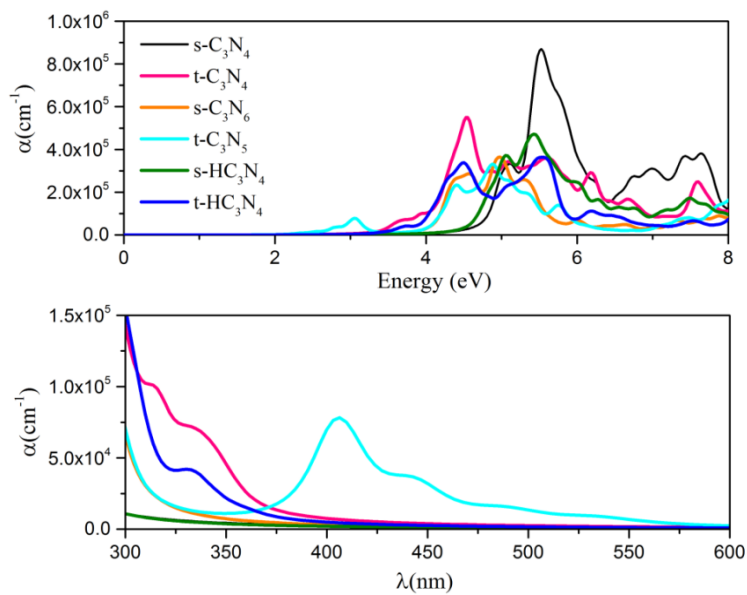

**Fig. 6,** Optical absorption spectra, α, of single-layer s-$C_3N_4$, t-$C_3N_4$, s-$C_3N_6$, t-$C_3N_5$, s-$HC_3N_4$ and t-$HC_3N_4$ as a function of photon energy (top panel) for the in-plane polarization, calculated using the RPA+HSE06 approach. Bottom panel shows a comparison of optical absorption spectra for the in-plane polarization as a function of wavelength, for aforementioned nanosheets in the visible range (350–600 nm) of light.

Our extensive theoretical results, confirm the high thermal stability and remarkable mechanical strength of s-$C_3N_6$, t-$C_3N_5$, s-$HC_3N_4$ and t-$HC_3N_4$ nanosheets. Moreover, the ultralow thermal conductivities along with direct-gap semiconducting characters of aforementioned nanomembranes are highly promising features to design novel carbon-based thermoelectric devices. Likely to original triazine-based graphitic carbon nitride nanosheets, highly porous atomic lattices, semiconducting electronic nature and appealing optical absorbance of these novel 2D systems propose them as attractive building blocks for the nanoelectronics, optoelectronics, nanosensors, biosensors, photocatalytic, electrocatalytic, photovoltaic, energy storage and adsorbent applications [28,73–76]. Nonetheless, the exploration of these application prospects requires more extensive theoretical and experimental studies.

## 4. Concluding remarks

Motivated by most recent experimental synthesis of $C_3N_5$ and $C_3N_4$ nanomembranes, and also by taking into account the well-known $C_3N_4$ triazine-based structures, in this work we predicted $C_3N_6$ and $C_3N_4$ as novel 2D materials. We then explored the mechanical properties, thermal stability, lattice thermal conductivity, optical adsorption and electronic features of experimentally realized and theoretically predicted porous carbon nitride nanosheets via extensive density functional theory calculations. Uniaxial tensile simulation results confirm elastic stability and highlight remarkable strength and acceptable stretchability of studied nanosheets, stemming from their strong covalent bonding networks. Ab-initio molecular dynamics trajectories confirm that despite of highly porous atomic lattices, all considered carbon nitride nanosheets could stay completely intact at the very high temperature of 1500 K. Non-equilibrium molecular dynamics results on the basis of machine learning interatomic potentials predict three orders of magnitude suppressed thermal conductivities for these novel carbon nitride nanomembranes in comparison with that of the graphene. HSE06 band structure calculations reveal that the considered monolayers are all semiconducting with (quasi-) direct band gaps ranging from 2.12 to 4.74 eV. The dependency of resulting electronic structure with respect to the core motifs and linkage groups are also explored. Our results showed that these novel 2D materials can absorb light in the ultraviolet and visible regions, appealing for optoelectronics and nanoelectronics applications. Extensive first-principles based results by this study provide a comprehensive vision on the stability, mechanical, thermal conduction, electronic and optical responses of $C_3N_4$, $C_3N_5$ and $C_3N_6$ as novel 2D semiconductors and suggest them as promising candidates for the design of advanced nanoelectronics, optoelectronics, nanosensors, biosensors, catalytic and energy storage/conversion systems.

Acknowledgment

B.M. and X.Z. appreciate the funding by the Deutsche Forschungsgemeinschaft (DFG, German Research Foundation) under Germany's Excellence Strategy within the Cluster of Excellence PhoenixD (EXC 2122, Project ID 390833453). A.V.S. was also supported by the Russian Science Foundation (Grant No 18-13-00479).

Appendix A. Supplementary data

The following are the supplementary data to this article:

Supporting Information

# Nanoporous $C_3N_4$, $C_3N_5$ and $C_3N_6$ nanosheets; Novel strong semiconductors with low thermal conductivities and appealing optical/electronic properties

Bohayra Mortazavi*[,a,b], Fazel Shojaei[c,d], Masoud Shahrokhi[e], Maryam Azizi[d], Timon Rabczuk[f], Alexander V. Shapeev[g] and Xiaoying Zhuang**[,a,f]

[a]Chair of Computational Science and Simulation Technology, Department of Mathematics and Physics, Leibniz Universität Hannover, Appelstraße 11,30157 Hannover, Germany.
[b]Cluster of Excellence PhoenixD (Photonics, Optics, and Engineering–Innovation Across Disciplines), Gottfried Wilhelm Leibniz Universität Hannover, Hannover, Germany.
[c]Department of Chemistry, College of Sciences, Persian Gulf University, Boushehr 75168, Iran.
[d]School of Nano Science, Institute for Research in Fundamental Sciences (IPM), Tehran, Iran.
[e]Department of Physics, Faculty of Science, University of Kurdistan, Sanandaj, Iran.
[f]College of Civil Engineering, Department of Geotechnical Engineering, Tongji University, Shanghai, China.
[g]Skolkovo Institute of Science and Technology, Skolkovo Innovation Center, Nobel St. 3, Moscow 143026, Russia.

*E-mail: bohayra.mortazavi@gmail.com

1. Atomic structures in VASP POSCAR format.

2- Phonon dispersion relations for the predicted t-H$C_3N_4$ and s-$C_3N_6$ monolayers.

3- AIMD results for the thermal stability.

4- Charge density distribution of VBM and CBM states.

1. Atomic structures in VASP POSCAR format.

s-C3N4
1.00000000000000
    4.7842289992586231    0.0000000000000000    0.0000000000000000
   -2.3921145001293280    4.1432638514578919    0.0000000000000000
    0.0000000000000000    0.0000000000000000   15.0000000000000000
  C    N
   3    4
Direct
 0.4901791484096898  0.9803579955387230  0.5000000007464354
 0.0196423692683693  0.5098211057673385  0.5000000000241522
 0.4901795040440148  0.5098202626505081  0.5000000011373595
 0.1701263316295368  0.8298738856915344  0.5000000023163407
 0.1701265063818060  0.3402525198699708  0.4999999936372895
 0.6597470668246924  0.8298731718163589  0.4999999989302001
 0.6666670734418921  0.3333330586655647  0.5000000032082155

t-C3N4
   1.00000000000000
    7.1340029738171848    0.0000000000000000    0.0000000000000000
   -3.5670014869085929    6.1782278061897946    0.0000000000000000
    0.0000000000000000    0.0000000000000000   15.0000000000000000
  C    N
   6    8
Direct
 0.7262151919359709  0.7128774668130546  0.5000000000000000
 0.0643908717306942  0.7127896543377616  0.5000000000000000
 0.4111810692046518  0.3862419656134648  0.5000000000000000
 0.7262516395821521  0.3747283371818639  0.5000000000000000
 0.0527807674016927  0.3861411483556836  0.5000000000000000
 0.0528344492571975  0.0278022291442852  0.5000000000000000
 0.1722628856618300  0.2667414947182536  0.5000000000000000
 0.5109291485576435  0.2703476125188189  0.5000000000000000
 0.8371212065311724  0.2702819738298174  0.5000000000000000
 0.5109039142709833  0.6019204585203539  0.5000000000000000
 0.8389497899222107  0.6001384245074917  0.5000000000000000
 0.1687055300275020  0.6018150233721684  0.5000000000000000
 0.8371903733670223  0.9281611939194931  0.5000000000000000
 0.1687731624244790  0.9281058107419327  0.5000000000000000

s-HC3N4
   1.00000000000000
     8.6474629750340810    0.0000000000000000    0.0000000000000000
     4.3237314870095309    7.4889226146832906    0.0000000000000000
     0.0000000000000000    0.0000000000000000   16.0000000000000000
   C    N    H
   6    9    3
Direct
  0.4124579021659329  0.1324211979905030  0.5000000000000000
  0.2139323951671415  0.4259617048373912  0.5000000000000000
  0.1189423969985839  0.2274667395765064  0.5000000000000000
  0.0445939123871995  0.7645903875504843  0.5000000000000000
  0.9495972997449371  0.0580518219303201  0.5000000000000000
  0.7511347988413490  0.9630107028987922  0.5000000000000000
  0.3881807451708994  0.2970029615110619  0.5000000000000000
  0.2835202209142693  0.0871424505321059  0.5000000000000000
  0.1926631893315616  0.5952723234011685  0.5000000000000000
  0.0736470527491306  0.4016783699581339  0.5000000000000000
  0.9709180810115683  0.2061241337792268  0.5000000000000000
  0.0899146483564124  0.8934881856322079  0.5000000000000000
  0.7754029293513156  0.1033209596330025  0.5000000000000000
  0.8800472681344331  0.7888017838869033  0.5000000000000000
  0.5817815450371171  0.9843598989743541  0.5000000000000000
  0.8532648888911041  0.3238195417683585  0.5000000000000000
  0.3103493508796191  0.5952753811388831  0.5000000000000000
  0.5817514778673996  0.8667013050005892  0.5000000000000000

s-C3N6
   1.00000000000000
    10.5746747409128208    0.0000000000000000    0.0000000000000000
     5.2873373704564131    9.1579369627302807    0.0000000000000000
     0.0000000000000000    0.0000000000000000   15.0000000000000000
   C    N
   6   12
Direct
  0.5664162472084939  0.1531107695198202  0.5000000000000000
  0.3383268475870338  0.3442715297475871  0.5000000000000000
  0.5295406715671476  0.3811593869523762  0.5000000000000000
  0.7595166985908435  0.5383201575005074  0.5000000000000000
  0.9512385591994246  0.5744160277363477  0.5000000000000000
  0.7234037219950835  0.7661087094093517  0.5000000000000000
  0.4228253055730207  0.1992695106930995  0.5000000000000000
  0.6269484164607846  0.2375928421320523  0.5000000000000000
  0.1844905226992708  0.3909960979690510  0.5000000000000000
  0.3845165402335411  0.4416811585382519  0.5000000000000000
  0.1050894432028926  0.5274163276790773  0.5000000000000000
  0.5761603369588784  0.4883583766612034  0.5000000000000000
  0.7125676207771718  0.4314206449861047  0.5000000000000000
  0.9046039857284995  0.4773754814954643  0.5000000000000000
  0.6624487295376014  0.6819746829731130  0.5000000000000000
  0.8670760593387286  0.7195039354716499  0.5000000000000000
  0.6165372550235162  0.9199232442217493  0.5000000000000000
  0.6735429863181110  0.9993312333131996  0.5000000000000000

```
t-HC3N4
   1.00000000000000
     12.9718286602075406    0.0000000000000000    0.0000000000000000
      6.4859143301037641   11.2339331561570397    0.0000000000000000
      0.0000000000000000    0.0000000000000000   16.0000000000000000
   C  N  H
   12  17   3
Direct
  0.5736432543870649  0.5497037548701602  0.5000000000000000
  0.5159412501479206  0.4101684825866895  0.5000000000000000
  0.7872460185400355  0.4777906072522372  0.5000000000000000
  0.7153066765215279  0.3361488318792127  0.5000000000000000
  0.2636815177947902  0.2396394001559773  0.5000000000000000
  0.1383116968102118  0.1656224862201094  0.5000000000000000
  0.3355754496826364  0.0259664579740179  0.5000000000000000
  0.9126458722281454  0.2784312038465869  0.5000000000000000
  0.4772174016813339  0.0979647718876603  0.5000000000000000
  0.5349418737916505  0.9006229829014638  0.5000000000000000
  0.6477000288404424  0.6750423249963483  0.5000000000000000
  0.4031863985882079  0.2973776915599329  0.5000000000000000
  0.4848639092330196  0.5263363469832614  0.5000000000000000
  0.6920651601939918  0.4545480001070135  0.5000000000000000
  0.5523422464309821  0.6615448754547725  0.5000000000000000
  0.7638569561783424  0.5899479231262834  0.5000000000000000
  0.6247941907477211  0.3148178623551705  0.5000000000000000
  0.8275097563138555  0.2473883182059520  0.5000000000000000
  0.0254731740707087  0.1795411945154932  0.5000000000000000
  0.2233863422687366  0.0494368784459176  0.5000000000000000
  0.4260816209853795  0.9141342986180572  0.5000000000000000
  0.1518212494418099  0.2609644970243468  0.5000000000000000
  0.3588148716204585  0.1211937164131013  0.5000000000000000
  0.8991001072165937  0.3872958984293845  0.5000000000000000
  0.2870341875034087  0.3284230074060381  0.5000000000000000
  0.6338308530786284  0.7878349358264212  0.5000000000000000
  0.5659714017492183  0.9857701460707951  0.5000000000000000
  0.4985322165178415  0.1885014073357064  0.5000000000000000
  0.4170471159689220  0.3962839907798704  0.5000000000000000
  0.0254867212041957  0.1009761077972461  0.5000000000000000
  0.3384840098909380  0.4748324326670320  0.5000000000000000
  0.7123864273712772  0.7878296703077402  0.5000000000000000
```

```
t-C3N5
   1.00000000000000
    15.1277409697215397    0.0000000000000000    0.0000000000000000
     7.5638704848607610   13.1010079769474004    0.0000000000000000
     0.0000000000000000    0.0000000000000000   15.0000000000000000
   C    N
   12   20
Direct
  0.4822644636641620  0.1085787326135701  0.5000000000000000
  0.3214275004720125  0.2440545548874637  0.5000000000000000
  0.4720142759863071  0.2636702855973905  0.5000000000000000
  0.1663292627727962  0.3888322519749068  0.5000000000000000
  0.3016889010575318  0.4143559340451146  0.5000000000000000
  0.4465543211214930  0.4245234721921383  0.5000000000000000
  0.6171240553036205  0.5231017381275944  0.5000000000000000
  0.7619102393974673  0.5334618991560535  0.5000000000000000
  0.8971530944588317  0.5591333373786449  0.5000000000000000
  0.7420006871770273  0.7038832251935508  0.5000000000000000
  0.5915388610112199  0.6840158883227199  0.5000000000000000
  0.5810802006627450  0.8392270826531424  0.5000000000000000
  0.5309358552134957  0.1616006457119609  0.5000000000000000
  0.3817503853957624  0.1431851514880575  0.5000000000000000
  0.2193696830181074  0.2871465032675500  0.5000000000000000
  0.3650419860246946  0.3073602638186950  0.5000000000000000
  0.5125092784839281  0.3240132461033267  0.5000000000000000
  0.0576086115771718  0.4261718604877913  0.5000000000000000
  0.2008522783995232  0.4548180502746642  0.5000000000000000
  0.3448574944995109  0.4732393208716329  0.5000000000000000
  0.0058670175868676  0.5218190496483875  0.5000000000000000
  0.4840246002400462  0.4958061838857145  0.5000000000000000
  0.5796813270672944  0.4517962752598863  0.5000000000000000
  0.7188101144715653  0.4744928347786221  0.5000000000000000
  0.8627191719608562  0.4930741013154951  0.5000000000000000
  0.5511099142572550  0.6236156329435260  0.5000000000000000
  0.6984814556736257  0.6404546973881480  0.5000000000000000
  0.8440585848798418  0.6608161495838836  0.5000000000000000
  0.5325177070693907  0.7861054650462402  0.5000000000000000
  0.6815883883005875  0.8047235903839933  0.5000000000000000
  0.5096863947248821  0.9480316867250476  0.5000000000000000
  0.5536037920703982  0.9997705598750607  0.5000000000000000
```

## 2- Phonon dispersion relations of the predicted t-HC$_3$N$_4$ and s-C$_3$N$_6$ monolayers.

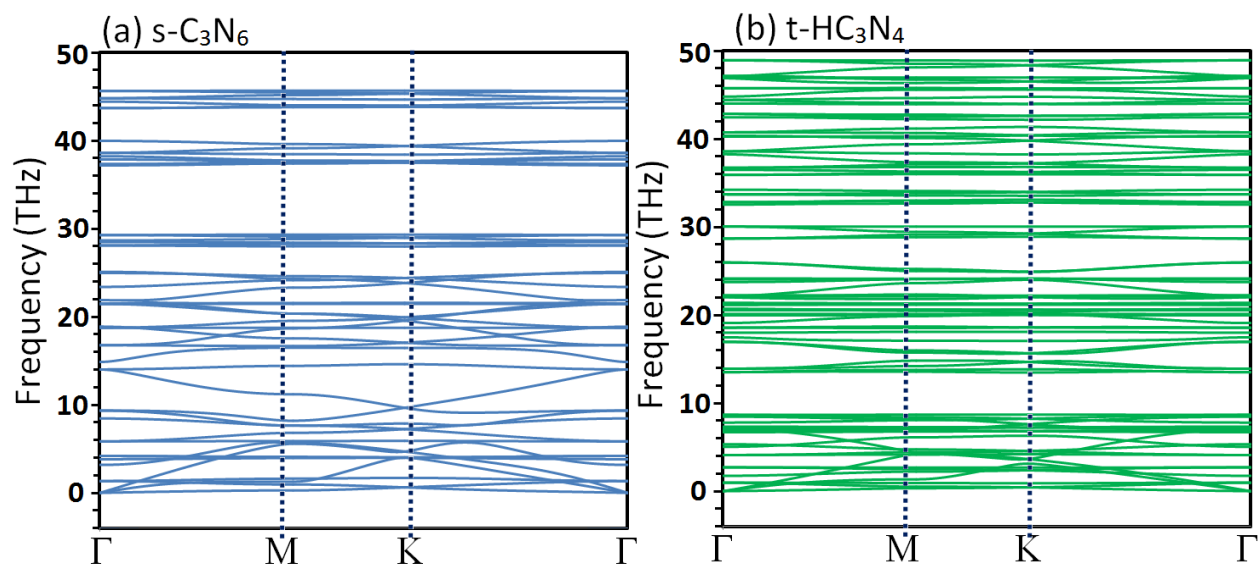

**Fig. S1**, Phonon dispersion relations of the predicted t-HC$_3$N$_4$ and s-C$_3$N$_6$ monolayers along high symmetry directions of the first Brillouin zone. For the t-HC$_3$N$_4$ monolayer flat bands appear in the phonon dispersion relations at frequencies around 104.3 THz, which are not shown here.

## 3- AIMD results for the thermal stability.

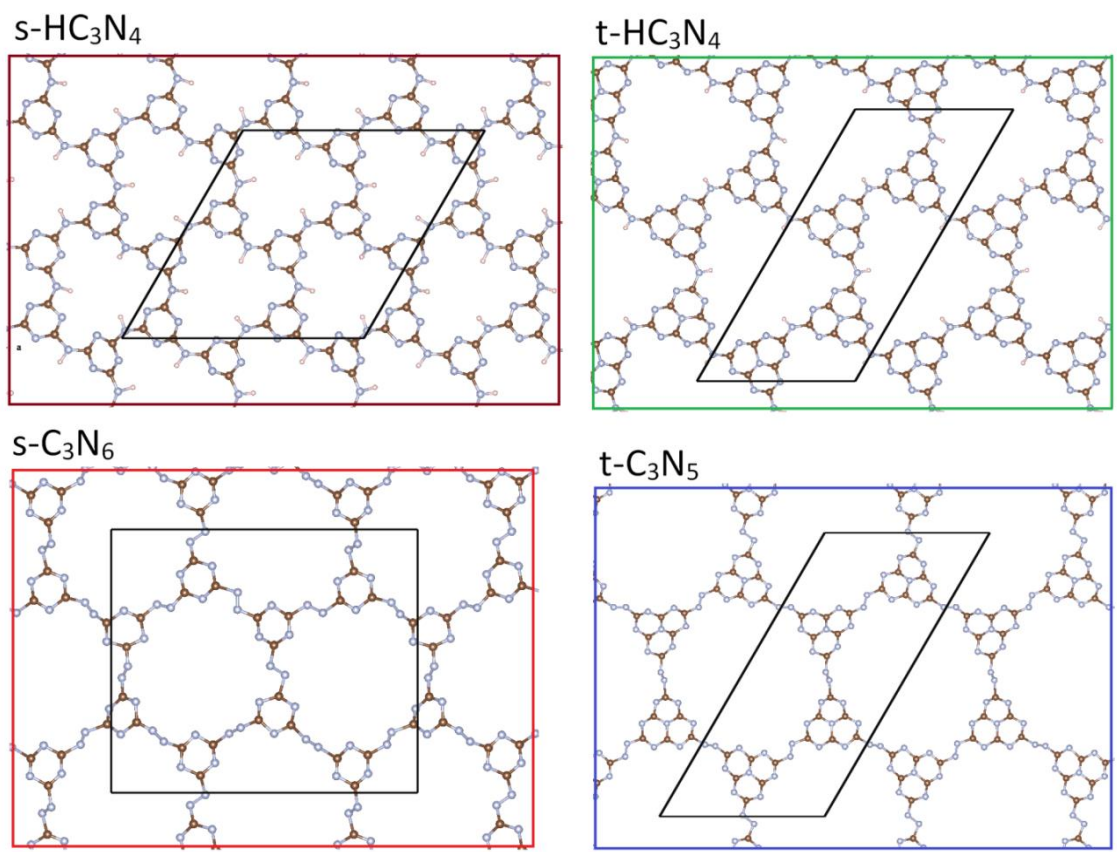

**Fig. S2**, Top views of carbon nitride monolayers after the AIMD simulations at 1500 K for 10 ps.

4- Charge density distribution of VBM and CBM states.

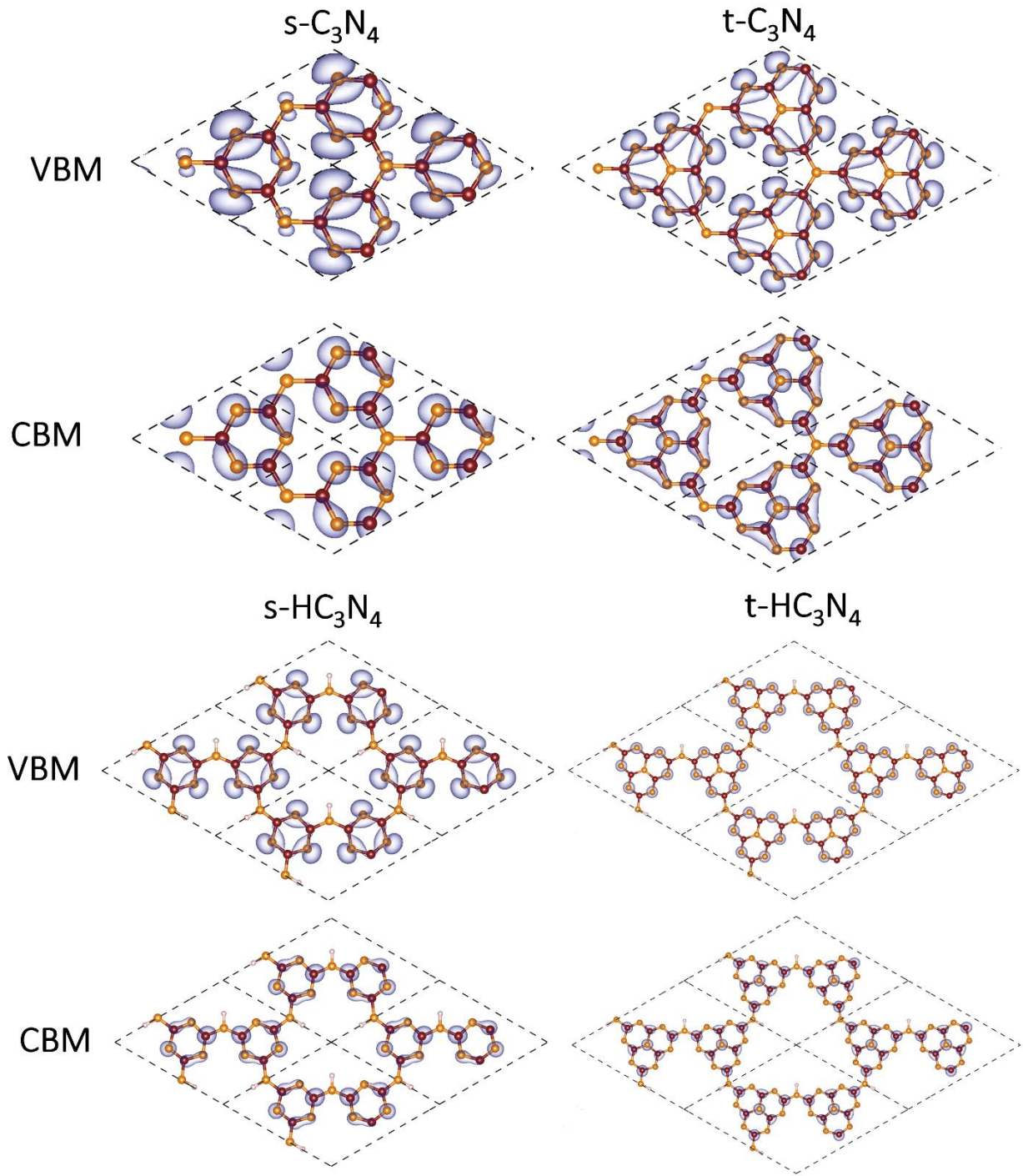

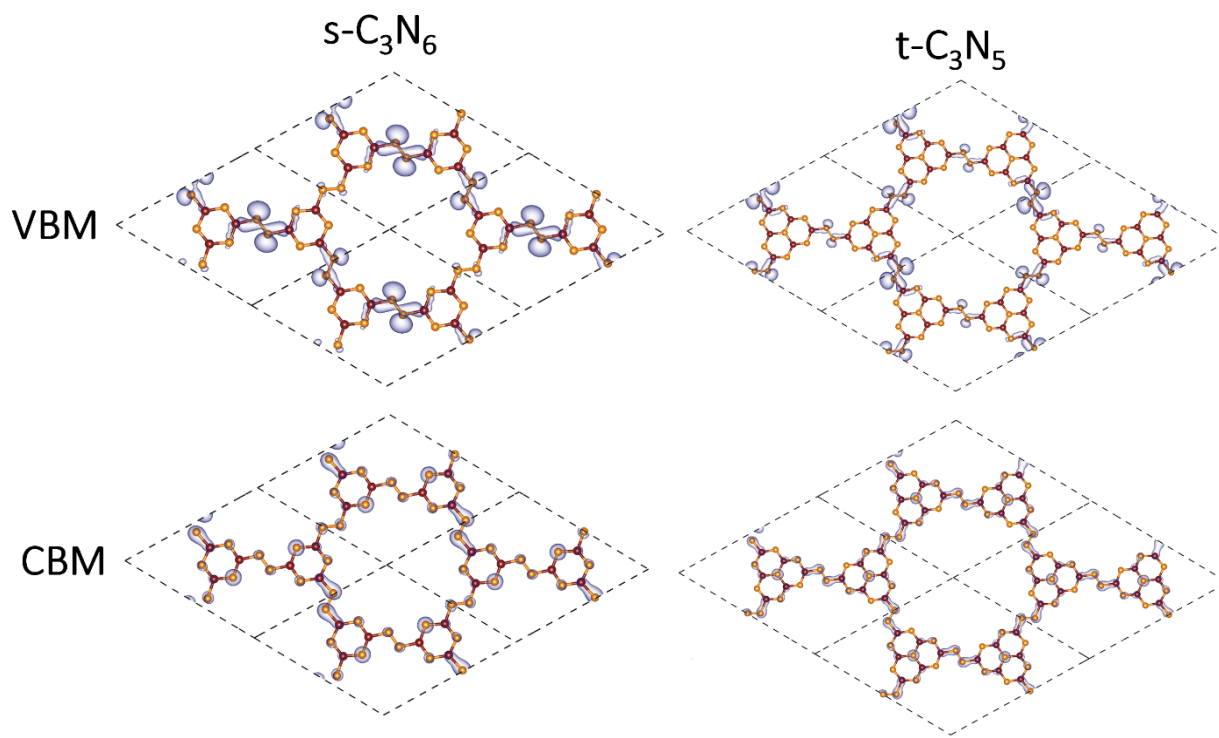

**Fig. S3**, Charge density distribution of VBM and CBM states of different carbon nitride monolayers we considered in this work. The iso-surface of electron density is set to 0.005 e/Å$^3$.